%% file: main.tex
\documentclass[letterpaper,twocolumn,10pt]{article}
\usepackage{main}

\input{preamble}
\input{macros}

\begin{document}

\date{}

\title{\name{}: Flexible Approximate Cache System for Video Diffusion}



\author{{\rm Desen Sun, Henry Tian, Tim Lu, and Sihang Liu}\\University of Waterloo}

\maketitle

\input{sec/0_abstract}
\input{sec/1_intro}
\input{sec/2_background}

\input{sec/3_highlevel}
\input{sec/4_impl}

\input{sec/5_eval}
\input{sec/7_disscussion}
\input{sec/8_conclusions}

\newpage

\bibliographystyle{plain}
\bibliography{bib/ml, bib/sys, bib/misc}

\end{document}

%% file: preamble.tex
\usepackage{subcaption}
\usepackage{amsmath}
\usepackage{xspace}
\usepackage{multicol}
\usepackage{multirow}
\usepackage{booktabs}
\usepackage{cleveref}
\usepackage[super]{nth}
\usepackage{enumitem}
\usepackage{amsmath}
\usepackage{svg}

\usepackage{siunitx}
\DeclareSIUnit\dollar{\$}
\DeclareSIUnit{\month}{month}
\DeclareSIUnit{\thousand}{k}
\DeclareSIUnit{\million}{M}

\usepackage{array}
\newcolumntype{L}[1]{>{\raggedright\let\newline\\\arraybackslash\hspace{0pt}}m{#1}}
\newcolumntype{C}[1]{>{\centering\let\newline\\\arraybackslash\hspace{0pt}}m{#1}}
\newcolumntype{R}[1]{>{\raggedleft\let\newline\\\arraybackslash\hspace{0pt}}m{#1}}

\usepackage{pifont}
\newcommand{\circled}[1]{\ding{\numexpr#1+201}}

\usepackage{pgfplots}
\usepgfplotslibrary{fillbetween}
\usepackage{tikz}
\usetikzlibrary{matrix,positioning}
\usetikzlibrary{external}
\usepackage{pgfplots}
\usepackage{pgfplotstable}
\usetikzlibrary{pgfplots.groupplots}
\usetikzlibrary{arrows}
\usetikzlibrary{patterns}
\usetikzlibrary{positioning}
\usetikzlibrary{decorations.pathreplacing}
\usetikzlibrary{shapes.arrows}
\usetikzlibrary{shapes.geometric,shapes.misc}
\usetikzlibrary{pgfplots.groupplots}
\pgfplotsset{compat=newest}
\pgfkeys{/pgf/number format/.cd,1000 sep={}}

\pgfplotsset{
    discard if/.style 2 args={
        x filter/.code={
            \edef\tempa{\thisrow{#1}}
            \edef\tempb{#2}
            \ifx\tempa\tempb
                
            \fi
        }
    },
    discard if not/.style 2 args={
        x filter/.code={
            \edef\tempa{\thisrow{#1}}
            \edef\tempb{#2}
            \ifx\tempa\tempb
            \else
                
            \fi
        }
    }
}

\makeatletter
\newcommand\resetstackedplots{%
\pgfplots@stacked@isfirstplottrue
}
\makeatother

\usetikzlibrary{spy}
\usetikzlibrary{patterns,backgrounds}
\pgfdeclarelayer{foreground}
\pgfsetlayers{background,main,foreground}

\usepgfplotslibrary{statistics}



%% file: macros.tex
\newcommand{\name}{FlexCache\xspace}
\newcommand{\cachename}{LRBU\xspace}

\newcommand{\ie}{i.e.,\xspace}

\newcommand{\nirvana}{NIRVANA\xspace}
\newcommand{\machinetype}{\texttt{a2-highgpu-1g}\xspace}
\newcommand{\gpucost}{\SI{3.67}{\dollar\per\hour}\xspace}
\newcommand{\disktype}{\texttt{balanced persistent disk}\xspace}
\newcommand{\storagecost}{\SI{0.1}{\dollar\per\giga\byte}\xspace} 

\DeclareMathOperator*{\minimize}{minimize}

%% file: sec/0_abstract.tex
\begin{abstract}
    Text-to-Video applications receive increasing attention from the public. Among these, diffusion models have emerged as the most prominent approach, offering impressive quality in visual content generation. However, it still suffers from substantial computational complexity, often requiring several minutes to generate a single video. While prior research has addressed the computational overhead in text-to-image diffusion models, the techniques developed are not directly suitable for video diffusion models due to the significantly larger cache requirements and enhanced computational demands associated with video generation. 

    We present \name{}, a flexible approximate cache system that addresses the challenges in two main designs. First, we compress the caches before saving them to storage. Our compression strategy can reduce 6.7$\times$ consumption on average. Then we find that the approximate cache system can achieve higher hit rate and computation savings by decoupling the object and background. We further design a tailored cache replacement policy to support the two techniques mentioned above better. Through our evaluation, \name{} reach 1.26$\times$ higher throughput and 25\,\% lower cost compared to the state-of-the-art diffusion approximate cache system.

\end{abstract}

%% file: sec/1_intro.tex
\section{Introduction}

Generative models have transformed content creation by enabling the generation of high-quality images or videos through user prompts. Diffusion models have emerged and attracted the most attention \cite{NEURIPS2021_49ad23d1, Ye_Liu_Wu_Wu_2024, 10655871, jeong2024groundavideo, 10656326, 10657475}. 
One of the prominent applications of these models is video generation, which has been extensively adopted in commercial platforms \cite{sora, kling, firely, make_a_video, Veo}. 
In the text-to-video generation, given a text prompt describing the content, a diffusion model starts with Gaussian noise and then continuously predicts and eliminates the noise over a number of denoising steps.
Such a process requires hundreds of convolution and transformer operations before generating the final video,  bringing a huge computation overhead and time cost.

A variety of studies have been aiming to improve the performance of diffusion models. For example, there have been parallelization approaches that divide the whole generation task into several patches and dispatch them to multiple GPUs \cite{xdit, distrifusion}.
Cambricon-D \cite{cambricon} quantizes the differential values from adjacent diffusion steps to an extremely low-bit representation, allowing for fast processing. 
Although these optimizations reduce the latency of video generation, the diffusion model is still compute-intensive and takes a long time to be processed on expensive GPUs. 
This is especially challenging for text-to-video diffusion models. 
For example, using the most popular text-to-image diffusion model Stable Diffusion-XL \cite{podell2023sdxl} to generate a high-quality image only takes an A100 GPU 8 seconds. In comparison, the text-to-video model VideoCrafter2 \cite{videocrafter} takes an A100 4 minutes to generate a high-quality seconds-long video.

Alternative to optimizing models, another approach is to start the generation from partially computed results rather than the initial Gaussian noise.
Recent work on text-to-image diffusion models, \nirvana{}, has proposed an approximate caching technique that saves the latent states (\ie intermediate results during computation) of a few key steps in a vector database \cite{nirvana}. 
Because of the step-by-step denoising algorithm, the latent state from a later step is closer to the final output. 
If an incoming prompt is similar to one of the previously cached prompts (\ie a cache hit), it loads its saved latent state as the starting point to process the new prompt and skip the cached steps. 
In addition, a higher prompt similarity allows a later step to be chosen. 
Text-to-video diffusion models also perform step-by-step denoising. 
However, we find that the existing caching approach does not directly apply to video generation.
We identify two main challenges that stem from the key differences between image and video generation. 

The first challenge comes from the cache size. 
Videos are much larger than images as they contain a sequence of frames.
The same capacity explosion applies to latent states --- the latent state cache of a 64-frame video takes $64\times$ higher capacity than an image under the same resolution.
Therefore, the same storage capacity can only cache latent states of much fewer video-generation prompts, leading to a lower hit rate. 
For example, according to \nirvana{}, a \SI{100}{\giga\byte} cache saves 18\% computations but it requires \SI{6.4}{\tera\byte} to achieve the same savings in video generation. 

The larger size of videos inherently results in significantly higher computational complexity. As illustrated in the aforementioned example, generating a 64-frame video takes around 30$\times$ longer than producing a single image of the same resolution.
In the case of lightweight caching mechanisms for image diffusion, like the one from \nirvana{}, are designed to maintain efficiency, as image generation typically takes only a few seconds.
However, for video diffusion models, the trade-off shifts towards optimizing caching strategies. Enhancements in cache design for video diffusion models can yield substantial savings, as any improvement in the cache hit rate directly reduces its computation overhead.
Consequently, the second challenge lies in designing an effective caching system to address the computational complexity of video diffusion models.

This work aims to enable efficient approximate caching for text-to-video diffusion models. 
Our system, \name{}, incorporates two core techniques aimed at reducing cache size and enhancing cache hit rates. 
Our analysis reveals that the inherent similarity within latent states offers significant potential for compression. 
We identify two types of similarities. 
First, inter-frame similarity is evident, where certain frames exhibit redundancy. Drawing inspiration from prior research on video processing \cite{Djelouah_2019_ICCV, Wu_2018_ECCV}, which identifies and deduplicates similar frames, we observe that analogous patterns exist in latent states, allowing the retention of only unique key frames.
Second, we observe that objects and their movements in the final video remain consistent throughout denoising steps. Consequently, the differential values between frames (\ie differences) effectively represent identical movements across all denoising steps, making them suitable for compression.
By leveraging these compression opportunities, \name{} integrates key frame compression to achieve 6.7$\times$ cache size reduction, with minimal impact on video quality. 

\begin{figure}
    \centering
    \includegraphics[width=1\linewidth]{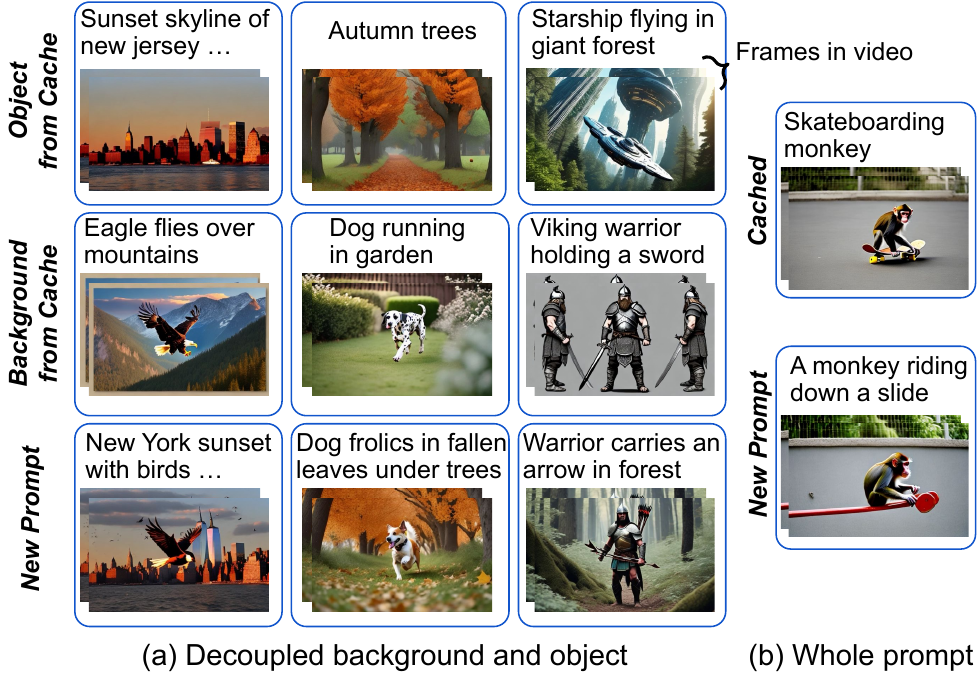}
    \caption{Example of \name{} generations.}
    \label{fig:intro_examples}
\end{figure}

To enhance the cache hit rate, we propose a caching mechanism specifically designed for text-to-video diffusion models. 
Instead of solely relying on the complete prompt for cache lookups, we find that decoupling object and background and performing separate lookups allow for 13.8\,\% higher hit rate. 
Building on this insight, \name{} incorporates text extraction along with object and background segmentation to reuse cached latent states from videos with the most similar object or background.
Depending on the similarity, \name{} dynamically selects between the whole prompt cache or combining object and background caches. 
Additionally, we design a novel cache replacement policy tailored to our compressed and decoupled caching scheme. 
It holistically accounts for the capacity of cached video latent states, cache access recency, and cache access frequency. 

We evaluate \name{} using VideoCrafter2 \cite{videocrafter} as the text-to-video diffusion model and VidProM \cite{wang2024vidprommillionscalerealpromptgallery} as the video prompt dataset, on a GCP instance with an A100 GPU. 
\Cref{fig:intro_examples}a and \ref{fig:intro_examples}b demonstrate video generations (the first frame) from the decoupled object and background caches and the whole prompt caches in \name{}, respectively. 
In both modes, \name{} achieves high quality. 
We summarize the contributions as follows:

\begin{itemize}[leftmargin=*,noitemsep,partopsep=2pt,topsep=2pt,parsep=0pt]
    \item We design a cache compression technique that leverages similarities in video diffusion models, allowing efficient latent state caching with almost no degradation in the quality. 
    \item We propose a cache lookup mechanism that decouples the object and background from the whole prompt to achieve a higher hit rate.
     We further design a new cache replacement policy tailored for video latent state caching that incorporates computations savings, cache size, and recency.
    \item We incorporate these techniques in \cachename{} that reuses latent states for text-to-video diffusion models. 
    \item We evaluate \name{} with a real-world dataset and a baseline that adapts \nirvana{}'s image caching to video cache. \name{} with a \SI{1000}{\giga\byte} cache storage improves the throughput by 26\,\% and saves 25\,\% cost on average over this baseline, with almost no degradation in video quality.
\end{itemize}

%% file: sec/2_background.tex
\section{Background and Motivation}

In this section, we first introduce diffusion models and approximate caching, and then present the challenges in optimizing video diffusion models. 


\subsection{Diffusion Model}
\begin{figure}[t]
  \begin{center}
  \includegraphics[width=\linewidth]{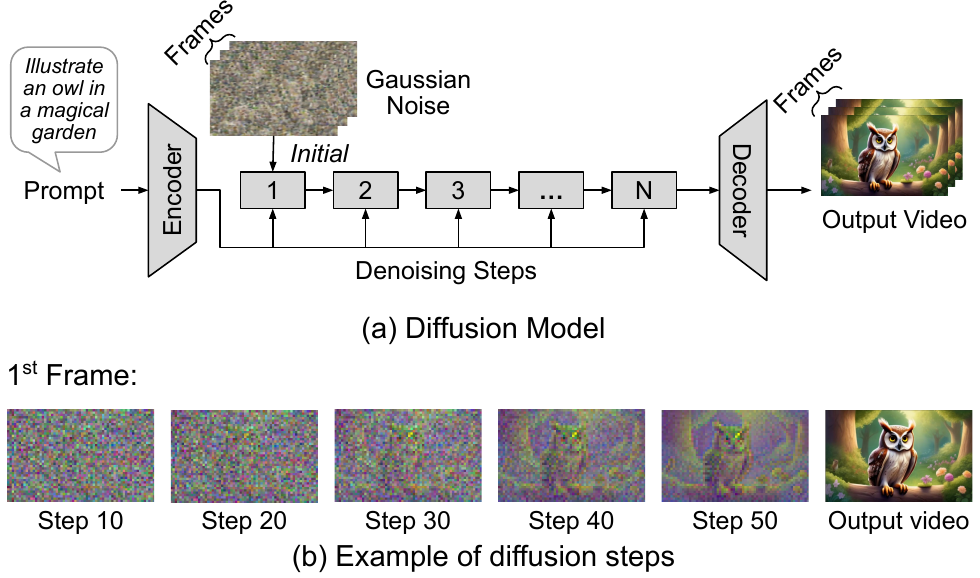}
  \end{center}
  \caption{\label{fig:diffusion} Diffusion model for video generation.
  }
\end{figure}

Diffusion models are a class of generative models that generate the image \cite{podell2023sdxl, ldm} or video \cite{videocrafter, svd} from prompts. 
Recent research has demonstrated that diffusion models achieve better performance than previous generation models on vision tasks \cite{NEURIPS2021_49ad23d1}. 
These models generate the output by taking a prompt and an initial Gaussian noise, and then eliminate unnecessary noise.
The denoise procedure takes multiple steps, making diffusion model computation intensive. Text-to-image diffusion and text-to-video diffusion have similar architectures, the primary difference is that image generation only applies one dimension transformer, while video tasks usually deploy spatial-temporal transformers \cite{wang2024videofactory, spatial_temporal_attention} to strengthen the interaction between spatial and temporal perceptions. 

\Cref{fig:diffusion}a illustrates the structure of a diffusion model. It takes Gaussian noise as the inital input and then goes through multiple denoising steps, where the output from the prior step is taken as the input to the next.
Each step processes the current noise and the embedding generated by the given prompt to predict the noise, and eliminate such noise from the input. 
After a number of steps, the noise is reduced and can be recovered to a high-quality video by decoder. The denoising part is the core component in diffusion models, typically based on U-Net \cite{ronneberger2015u, svd, videocrafter} or Transformer (DiT) \cite{dit, Chen_2024_CVPR, sora}.
\Cref{fig:diffusion}b demonstrates this process in video generation, where the noise gradually reduces and eventually becomes the final video over 50 denoising steps.
To alleviate the heavy computation complexity, previous research finds a way to infer diffusion model in latent space \cite{ldm}. We can see that although the quality increasingly gets better with step growth, the position of that owl never moves within the same frame. 


Diffusion models are costly as they take multiple steps to get the final output. 
There have been optimizations on the sampler to decrease the number of steps in diffusion models, such as decreasing the number of steps to 100 \cite{NEURIPS2020_4c5bcfec}, 50 \cite{song2021denoising}, and 20 \cite{lu2023dpmsolver}. These steps need to be processed in sequence, which occupies the majority overhead of diffusion models.
There have also been works that split the input to multiple patches \cite{xdit, distrifusion}, or schedule different modules in diffusion such as LoRA \cite{hu2022lora}, ControlNet \cite{controlnet} to suitable devices \cite{swiftdiffusion}, to achieve higher parallelism within each step,  reducing the latency. Although these proposals improve performance, the root problem of high compute complexity remains.

\begin{figure}[t]
  \begin{center}
  \includegraphics[width=\linewidth]{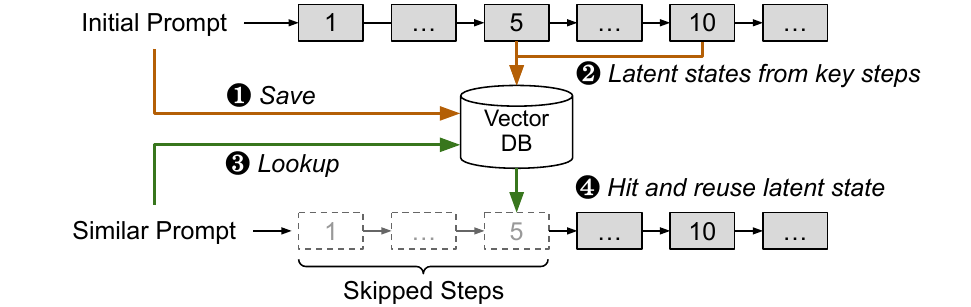}
  \end{center}
  \caption{\label{fig:prompt-cache} Approximate-caching for diffusion models.
  }
\end{figure}

\subsection{Caching for Diffusion Models}

To relieve the costly computation overhead, one solution is to store the latent states (\ie intermediate states during step-by-step computation) in a cache and load them to skip steps when a similar prompt arrives \cite{MLSYS2024_a66caa17, nirvana}. 
The similarity between the incoming prompt and the cached prompt determines the computations that can be saved. 

\nirvana{} \cite{nirvana} is the state-of-the-art work that adopts approximate cache to reduce computation in text-to-image diffusion models. 
As illustrated in \Cref{fig:prompt-cache}, upon a new request, \nirvana{} saves the prompt \circled{1} and latent states \circled{2} of key steps (\nth{5}, \nth{10}, \nth{15}, \nth{20}, and \nth{25}) in a cache. 
The cache is based on a vector database to enable lookup. 
Given a new prompt, \nirvana{} attempts to look up \circled{3} the most similar previous prompt using cosine similarity as the metric. 
If a cache is hit, the current request can set this cache as input \circled{4} (the \nth{5} step in this example) and proceed from the next step. 
The higher the prompt similarity score is, the more steps the incoming prompt can skip.
\nirvana{} has shown that the cached prompt can save up to 25 steps out of a total of 50 steps. 
Because the remaining steps continue to process the latent state, the output quality remains good as the diffusion model takes additional steps to continue the update. 

Caching is a promising solution to reduce computation cost, as \nirvana{} has demonstrated in image generation.
However, caching is not a one-size-fits-all solution to diffusion-based generation.
The process of generating videos via diffusion models is similar to that of images.
Straightforwardly adapting approximate caching, nonetheless, cannot reap the same benefit. 
Next, we describe the challenges in enabling caching for video generation. 


\subsection{Challenges in Caching Video Generation} \label{sec:challenges}

\begin{figure}[t]
  \begin{center}
  \input{results/computation_saving_by_memory}
  \end{center}
  \caption{\label{fig:memory-demand} Cache size required to achieve different levels of computation savings, for both image and video generations.}
\end{figure}
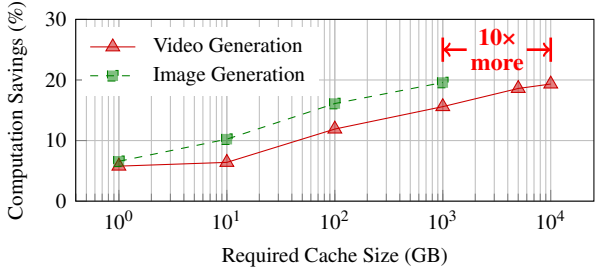

Cache hit rate and similarity scores are the two key aspects of approximate caching --- a high cache hit rate enables a larger number of prompts to reuse caches, while a high similarity score allows prompts to skip additional steps.
However, achieving a high hit rate and a high similarity score simultaneously is challenging. 
Unlike image generation models, video generation models feature substantially larger latent state sizes and higher computational complexity. 
We next explain these key differences in detail.



\textbf{Larger latent states.}
Typical video generation applications allow users to generate seconds-long videos \cite{sora, kling}, which consist of tens to hundreds of frames.
For example, one 64-frame video of resolution $320 \times 512$ with each pixel defined by 4 channels (RGBA) has a total size of $320 \times 512 \times 4$. 
The latent state is downsampled by $8\times$ but still has a size of  $40 \times 64 \times 4$.
This dimension leads to \SI{2.5}{\mega\byte} of latent state size.
Directly adapting the caching scheme in \nirvana{} takes a total capacity of \SI{12.5}{\mega\byte} to save the five key steps (\nth{5}, \nth{10}, \nth{15}, \nth{20}, and \nth{25}) of each prompt. 
In contrast, images are static, consisting of only a single ``frame''. Thus, caching the latent states of the image with the same dimension only requires \SI{200}{\kilo\byte}. 
We demonstrate the capacity challenge using the VidProM dataset \cite{wang2024vidprommillionscalerealpromptgallery} that comes from real user prompts (more methodology details in \Cref{subsec:dataset}). 
We evaluate an experiment with infinite storage and present the certain computation savings (y-axis) and the corresponding cache requirements (x-axis) in \Cref{fig:memory-demand}. 
To achieve 19.3\,\% computation savings, the cache capacity requirement is \SI{10}{\tera\byte}. 
Theoretically, generating images using identical prompts would require 10$\times$ less cache compared to video generation when employing the same resolution as \nirvana{}---significantly higher than videos.
As video diffusion models become increasingly widespread, we expect the capacity demand to exacerbate as models scale.


\textbf{Higher computational complexity.}
As one video consists of a sequence of frames, each diffusion step takes a more complex procedure than image generation. Different from image generation which only applies 2D convolution and spatial transformer, video generation adds another dimension --- temporal dimension for both convolution and transformer operations. 
As a result, using VideoCrafter2 to generate a 64-frame video of $320 \times 512$ resolution takes 242 seconds on an A100 GPU, whereas generating an image of the same resolution only takes 8 seconds using the Stable Diffusion-XL model \cite{podell2023sdxl}.
On the other hand, savings from caching allow the diffusion model to skip steps, which proportionally reduces the total generation time. For example, skipping 10 steps means 48 seconds for video generation while it only translates to 1.6 seconds for image generation. 
This stark difference in time scale indicates that the tradeoffs between caching performance and the complexity of the cache system have been reversed. 
In image diffusion caching, a lightweight caching lookup system is required as generating a single image only takes a matter of seconds. 
However, the high potential benefit in absolute time savings of video diffusion models motivates further optimizations in caching design for a higher hit rate and similarity score.


These challenges highlight the need for a more efficient caching scheme for text-to-video diffusion models. 
This work aims to redesign approximate caching by overcoming challenges that stem from the high latent size and high computational complexity. 
Next, we will present the high-level ideas.

%% file: results/computation_saving_by_memory.tex
\begin{tikzpicture}
\begin{axis}[
style={font=\footnotesize},
xlabel={Required Cache Size (GB)},
ylabel={Computation Savings (\%)},
width=1\hsize,
scaled y ticks=false,
xtick pos=bottom,
ytick pos=left,
xmode=log,
ytick = {0,10,20,30},
ymin=0,
ymax=30,
clip=false,
height=4cm,
grid=both,
legend columns=1,
legend cell align=left,
legend style={
cells={align=left},
anchor=north,
at={(0.24,0.96)},
draw=none,
column sep=1ex,
},
]
\addplot+[
draw=red!80!black,
mark=triangle*,
mark options={scale=1.5, fill=red!80!black, fill opacity=0.5}
] table[col sep=comma, x=memory size, y=computation savings] {results/csv/computation_saving_by_memory/vid_uncompressed.csv};
\addplot+[
dashed,
draw=green!50!black,
mark=square*,
mark options={scale=1, fill=green!50!black, fill opacity=0.5}
] table[col sep=comma, x=memory size, y=computation savings] {results/csv/computation_saving_by_memory/img.csv};
\legend{Video Generation,Image Generation};

\draw[draw=red, line width=0.35mm] (1000,23) -- (1000,27);
\draw[draw=red, line width=0.35mm] (10000,23) -- (10000,27);
\draw[->, draw=red, line width=0.35mm] (1666,25) -- (1000,25);
\draw[->, draw=red, line width=0.35mm] (6000,25) -- (10000,25);
\node[draw=none,align=center,text width=0.7cm,text=red] at (3200,25) {\small\bf 10\texttimes \\ more};

\end{axis}
\end{tikzpicture}

%% file: sec/3_highlevel.tex
\section{Overview of \name{}}

To overcome the aforementioned challenges, we proposed \name{} to achieve efficient caching for video diffusion models. 
We start with describing the high-level ideas and then present an overview of \name{}.

\subsection{High-level Ideas}

\subsubsection{Cache Compression}
\label{subsubsec:cache_compression}
\begin{figure}
\includegraphics[width=1\linewidth]{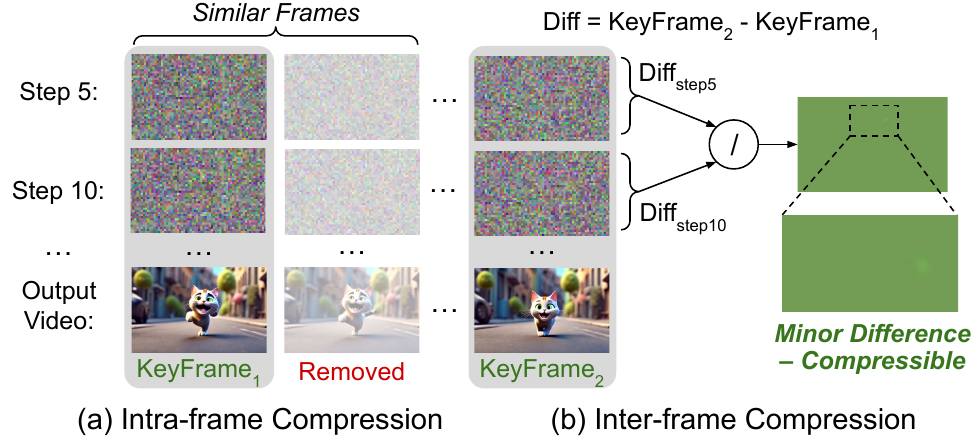}
\caption{\label{fig:compression_example} Example of intra-step and inter-step compression.}
\end{figure}

The large size of the latent cache is due to video frames. 
Nonetheless, the inherent similarity in latent states provides an opportunity to compress the cache size, and save more latent states in the cache. 
We observe two types of similarities. 
The first type is the similarity among frames. 
Some works point out that there are redundant frames in a video, allowing them to discard similar frames  \cite{Djelouah_2019_ICCV, Wu_2018_ECCV}. They inspire us that we can try to leverage the frame similarities in latent noise. Changes in a video do not happen suddenly but rather gradually.
Therefore, repeated frames can be removed to reduce the cache size, without degrading the quality. And, only a few \emph{key frames} need to be preserved. 
\Cref{fig:compression_example}a is an example video of a ``walking cat'', where the first two frames are almost identical and the second one can be removed from the cache. 
As diffusion models gradually reduce noise over steps, the same similarity among frames also exists across latent states. 
This type of similarity enables compression within each step. We refer to this compression technique as \emph{intra-step compression}.

The second type of similarity is the differential value similarity.
We find that even though diffusion models reduce the noise of latent states over steps (as demonstrated in \Cref{fig:diffusion}), the same frame among steps has the same content, following the same movement. 
In comparison, key frames within the same step have major changes as the scene of the video changes. 
Following this observation, if we obtain the difference between two key frames (\ie calculate their differential value) and compare them among steps, there should be minimum differences. \Cref{fig:differential} presents the average similarity scores for these differential values from \SI{5}{\thousand} videos.
It is clear that the differential values are almost identical within the first 25 steps, \ie the steps to be cached. 
Although the differential tends to be more dissimilar with step growth, it's useless to store the later several latent states according to \nirvana{}. 
\Cref{fig:compression_example}b demonstrates this observation. We calculate the differential ($\mathit{Diff}$) among $\mathit{KeyFrame1}$ and $\mathit{KeyFrame2}$ for both 5\textsuperscript{th} step and 10\textsuperscript{th} step, \ie $\mathit{Diff}_{\mathit{step}5}$ and $\mathit{Diff}_{\mathit{step}10}$, respectively. 
Then, we calculate the ratio between $\mathit{Diff}_{\mathit{step}5}$ and $\mathit{Diff}_{\mathit{step}10}$, ending up with an image with almost all pixels identical, except for a few pixels as shown in the zoom-in view. 
The ratio between differential frames can therefore be represented with a single floating point value, eliminating the need for saving all key frames. 
We refer to the second compression technique as \emph{inter-step compression}.
 
Together, the two compression techniques can reduce the video latent cache size by an average of $6.7\times$, with almost no degradation in the video quality. 
We explain the details about the cache compression in \Cref{subsec:compression}.




\subsubsection{Cache Hit Rate}

\begin{figure}
\begin{minipage}[t]{0.45\linewidth} 
  \centering
  \input{results/diff_heatmap}
  \caption{\label{fig:differential} Differential value similarity among steps.  }
\end{minipage}
\hspace{1mm}
\begin{minipage}[t]{0.52\linewidth} 
  \centering
  \input{results/similarity_hit}
  \caption{\label{fig:similarity-and-hit-rate} Whole prompt, object, and background hit rates.}
\end{minipage}
\end{figure}

Previous studies, such as \nirvana{}, compare the whole prompt's similarity with others, and choose the most similar existing prompt's cache as the input. It's always too strict to find another prompt that describes exactly the same thing as the current prompt. 
However, we found that the prompts of video generation requests always consist of two parts: the background and the object. This observation enables us to decouple the cache lookup. 

\Cref{fig:similarity-and-hit-rate} shows the distribution of prompts that have certain similarity scores. 
We assume that there is a pool of \SI{100}{\thousand} requests that have been served
and then we track the cosine similarity for the next \SI{1}{\thousand} prompts with the existing \SI{100}{\thousand} prompts.
We use a transformer-based text extractor to help us extract the key contents of the background and object from a given prompt. 
For background and foreground sentences, we also convert them to CLIP embeddings \cite{clip} and get their similarity scores. Based on \nirvana{}, the state-of-the-art approximate cache research in the diffusion model, requests can reuse the intermediate states only when their similarity score is over 0.65. The higher their similarity score is, the more computations the request can save. By decoupling the object and background from the whole prompt, the cache system achieves at least 13.8\,\% higher hit rate. 

Finally, we incorporate a novel cache replacement policy in our caching system. This policy takes account of both memory and recency which is ignored by \nirvana{}. Besides, unlike \nirvana{}, which only inserts cache when cache misses, we attempt to insert the cache no matter if it skips steps.

\subsection{System Overview} \label{subsec:sys_overview}

\begin{figure}[t]
  \begin{center}
  \includegraphics[width=\linewidth]{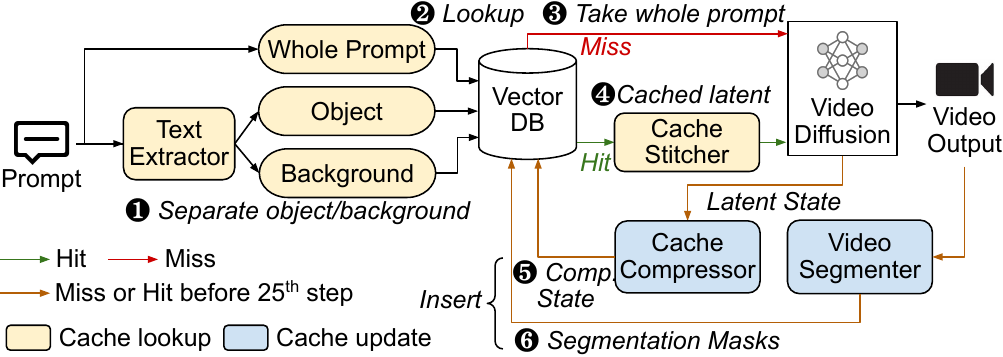}
  \end{center}
  \caption{\label{fig:arch} Overview of \name{}.}
\end{figure}

The cache compression and hit rate optimization aim to save more steps in video generation. Putting these ideas together, \Cref{fig:arch} shows the overview of \name{}. 
We divide the process into two phases: cache lookup and cache update. 

\paragraph{Cache lookup.}
When a new request arrives, a \emph{text extractor} separates the object and background description \circled{1}. 
Then, \name{} converts both the extracted prompt (\emph{object + background}) and the \emph{whole prompt} into embeddings using a CLIP model \cite{clip} and looks them up in a vector database (Vector~DB) \circled{2}. 
The Vector DB caches latent states in 5 key steps: \nth{5}, \nth{10}, \nth{15}, \nth{20}, and \nth{25}, similar to \nirvana{}'s strategy; it maintains three indices for lookup, corresponding to the whole prompt, the object part, and the background part. All indexing structures share the same cached latent states.  
The lookup is a hit if there is at least a pair of object + background or a whole prompt cache that is similar to the incoming prompt (similarity threshold is 0.65, following \nirvana{}); otherwise, it is a miss. 
Note that having only the object or background pass the similarity threshold is not regarded as a hit because video generation requires both. 
\name{}'s cache selection prioritizes the one with the highest similarity with the incoming prompt. We discuss the selection details in \Cref{subsubsec:cache_lookup}.
In the case of a \emph{miss}~\circled{3}, the video diffusion model processes the original prompt; in the case of a \emph{hit}~\circled{4},
the diffusion model starts with the cached latent state and thus takes fewer denoising steps.
A \emph{cache stitcher} combines the decoupled object and background caches when they have better similarity.

\paragraph{Cache update.}
When the video has been generated, \name{} saves its latent state back to the Vector DB if the prompt was a miss or the cached step was before the \nth{25} step (indicating a need for a better cache to serve future prompts that akin). If \name{} generates the video from the cache, then the latent states after the initial input (the reused cache) will be saved.
A \emph{cache compressor} compresses the states~\circled{5} (details in \Cref{subsec:compression}). 
In parallel, a \emph{video segmenter} generates a segmentation mask for each frame of the output video, depicting the boundary between the object and the background~\circled{6} (details in \Cref{subsubsec:cache_reconstruction}). 
Finally, \name{} inserts both the compressed latent states and the segmentation masks into the Vector DB for future prompts to look up.

\begin{figure}[t]
  \centering
  \includegraphics[width=1\linewidth]{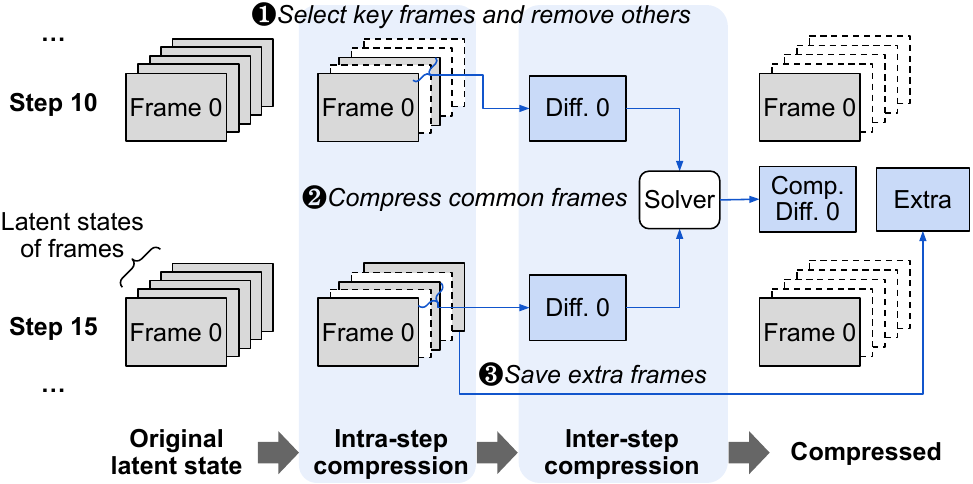}
  \caption{\label{fig:compression} Overview of Cache Compressor.
  }
\end{figure}

\section{Design of \name{}} \label{sec:design}

In this section, we describe the design of \name{} in detail, including the cache compression mechanism and optimization strategies that enable a higher hit rate. 

\subsection{Cache Compression} \label{subsec:compression}

We use a combination of intra- and inter-step compression techniques to reduce the size of cached latent states, allowing for more cached saved and achieving a higher hit rate.
\Cref{fig:compression} illustrates the workflow of this compression technique. Assuming that there are two latent states and each latent state has five frames. In the intra-step compression stage, we select and save only the key frames \circled{1}. 
A total of five frames that are similar to others are discarded after this stage.
Then, inter-step compression further reduces the size of key frames' latent states with a solver that reduces differential values of the common key frames \circled{2}. 
Note that not all the latent states have the same number of key frames, so there can be extra frames after inter-step compression. We save them as extra frames~\circled{3}. 
Eventually, the original ten frames are reduced to only four. 
Next, we describe both techniques in detail. 

\subsubsection{Intra-step Compression}
\label{subsubsec:intra-compression}

    

  

We first propose a latent state compression mechanism by leveraging the frame similarity. Unlike ground truth video compression, which needs to maintain the same quality after decompression, such latent states are still supposed to be processed for multiple steps, allowing for a certain degree of accuracy loss.
So it's possible to only save the key frames and simply repeat these frames when recovering.

We try to traverse the whole latent state from the last frame to the first frame when compressing. We get the similarity scores with the current frame and all the frames in front of it. Then we set a threshold to ensure that all of the other frames share enough similarities with the key frames. The detail of the threshold will be discussed in \Cref{sec:impl}
Next, we record the map relationship of the key frame and the compressed frame. If there are no other similar frames, then the current frame is also one key frame and it will map to itself. Finally, the original frame will be removed if it is not a key frame. In this way, we eliminate all of the unnecessary frames and only save the key frames and the relationship with the key frames, which take much less storage than the whole latent state. \Cref{subfig:redundant_frames} shows the average percentage of redundant frames. With steps growing, the frames become less similar since there are more details within frames.
When decompression, we can simply traverse the map and repeat the key frames to other positions. Although it is not exactly the same as the original latent state, the accuracy loss is acceptable because there are still some computations, which will not hurt the final quality. 
We will show the detailed evaluation in \Cref{subsubsec:compression_efficiency_quality}. 

\subsubsection{Inter-step Compression}
\label{subsubsec:inter-compression}

We further design a compression mechanism based on the observation that the differential values among steps are also similar.
For each cached denoising step $s$, the differential values of the m\textsuperscript{th} frame are its difference from the first frame, \ie $ \mathit{Diff}_{s}^{m} = \mathit{Frame}_{s}^{m} - \mathit{Frame}_{s}^{0}$. 
Based on the insight in \Cref{subsubsec:cache_compression}, it is possible to use the set of differential values of all key frames from one step (which we refer to as the \emph{base} $\mathit{Diff}_{\mathit{base}}$) to calculate those from other steps, effectively reducing the storage size.  
Note that some steps may have more key frames, which will be separately saved as extra frames in the cache.
The value that will be used to generate other differential values from the \emph{base} is defined as $\alpha_s$.
As there are 5 cached steps in \name{}, we have a total of 5 candidates for $\mathit{Diff}_{\mathit{base}}$.
The inter-step compression mechanism will attempt each of them and choose the one that leads to the highest similarity after decompression. 
We next explain the calculation for $\alpha_s$. 

We further denote $\mathit{i,j}$ as the i\textsuperscript{th} row and j\textsuperscript{th} column pixel in a frame, and use $\mathit{Diff}_{s}$ to collectively represent each key frame $\mathit{Diff}_{s}^{m}$ within step $s$.
Because $\mathit{Diff}_{s}$ and $\mathit{Diff}_{\mathit{base}}$ are known before compression, we model this as an inverse problem that infers model parameters from relevant observations.
Here we treat the $\mathit{Diff}_{s}$ and $\mathit{Diff}_{base}$ as the measurements and $\alpha_{s}$ is the parameter that needs to be inferred.
The inverse problem can be formulated as finding a solution to the operator equation $K(u) = f$, where $K$ is the forward operator, $u$ is the existing frame and $f$ is the measurement.
We therefore can define our equation as:
\begin{equation} \label{eqa:reverse}
K\left(\mathit{Diff}_{\mathit{base}}\right) = \mathit{Diff}_{s}
\end{equation}
As \Cref{fig:compression_example} and \ref{fig:differential} indicate the relationship between multiple differential values follows a linear-like pattern, the goal is to find the coefficient $\alpha_s$ that minimizes the average difference between $\alpha_s \times \mathit{Diff}_{\mathit{base}}$ and $\mathit{Diff}_{s}$.
Thus the inverse goal is to minimize  $\sum_{\mathit{i,j}}^{\mathit{Diff^m}}\left( \mathit{Diff}_{s}^{m,i,j} - \mathit{Diff}_{\mathit{base}}^{m,i,j}  \times\alpha_s\right)^{2}$.
This minimization is a standard least squares problem. We take the derivative of $\alpha_s$ and calculate its value when the derivative equals 0. Finally, we derive this forward operator:
\begin{equation} \label{eqa:derivative}
    K\left(\mathit{Diff}_{base}^m \right) = \frac{\sum_{\mathit{i,j}}^{\mathit{Diff^m}} \left(\mathit{Diff}_{s}^{m,i,j}  \times \mathit{Diff}_{base}^{m,i,j}\right) }{\sum_{\mathit{i,j}}^{\mathit{Diff^m}} \left(\mathit{Diff}_{base}^{m,i,j} \right)^2 } \times \mathit{Diff}_{base}^m
\end{equation}
The coefficient of the forward operator can be represented in a single value that occupies much less memory. In this way, we leverage the similarity across denoising steps to further compress the cache.

\subsection{Cache Hit Rate Optimizations} \label{subsec:hit_opt}
\begin{figure}
    \centering
    \includegraphics[width=1\linewidth]{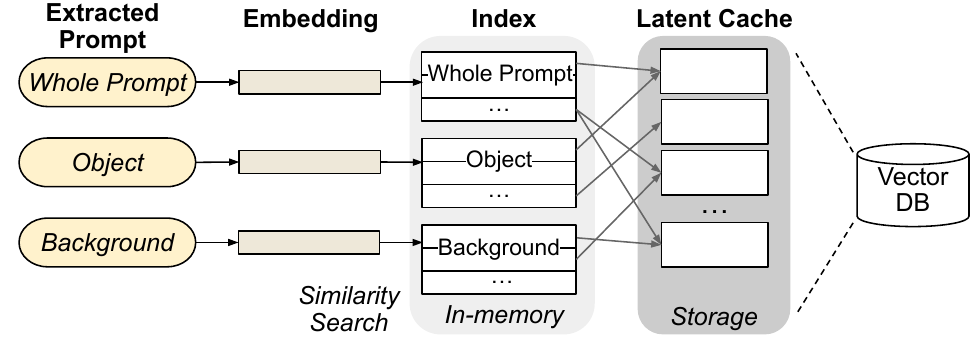}
    \caption{Vector database for cache lookup.}
    \label{fig:vdb_structure}
\end{figure}

In this section, we will describe our cache design that optimizes for cache hit rate through decoupled object and background lookup and latent reconstruction, and a tailored cache replacement policy.

\subsubsection{Cache Lookup} 
\label{subsubsec:cache_lookup}

We find that both prompt \cite{extract, Sun2024UMIE} and the final video \cite{background_sub} can be decoupled into two parts: the object that usually shows the main character or the event, the background that usually describes the general environment or settings. 
%

Based on this insight, we propose an object-background segmentation technique to achieve a higher hit rate and more computation savings. 
\Cref{fig:vdb_structure} demonstrates our design. 
For an incoming prompt, a text extractor first extracts the descriptions of the object and the background from the original prompt.
The vector DB maintains three index tables to look up the three kinds of prompts. 
The index tables are not memory-consuming and thus are kept in memory but the latent states are saved in storage. 
Next, \name{} looks up the original prompt and the extracted object and background prompts in the vector DB to find their most similar prompts, and get 3 similarity scores: $\mathit{Sim}_{\mathit{background}}, \mathit{Sim}_{\mathit{object}}, and \mathit{Sim}_{\mathit{whole}}$. 
Then it chooses the higher value from $\mathit{Sim}_{\mathit{whole}}$ and $\min(\mathit{Sim}_{\mathit{background}}, \mathit{Sim}_{\mathit{object}})$, to skip more steps.
When $\min(\mathit{Sim}_{\mathit{background}}, \mathit{Sim}_{\mathit{object}}) > \mathit{Sim}_{\mathit{whole}}$, applying the background-object segmentation will yield more computation savings. 
We show the fraction of hits from the whole prompt, and decoupled object and background prompts in \Cref{subsec:throughput}.

\subsubsection{Video Latent Reconstruction}
\label{subsubsec:cache_reconstruction}

When \name{} finds latent caches with higher similarity scores with the decoupled object and background descriptions than the whole prompt, the cache returns the latent states of the two videos: one provides the object and another provides the background. 
Next, \name{} needs to stitch these two caches into one as the input to the video diffusion model.

Simply getting the average value of these two latent states does not work well because both frames contain unneeded pixels. 
To precisely identify which part of the videos will be used as the object/background of the new generation, we adopt a segmentation model that takes the extracted object and background descriptions. 
However, latent states, especially those from the first several denoising steps, are mostly noise (as \Cref{fig:diffusion} shows), so the segmentation model can hardly detect any objects. 
However, the positions of these objects are never changed as the denoising steps increase.
Our approach is to segment only the final video by generating masks for objects and backgrounds, and then apply these masks to the previous latent states of the cached steps.
\name{} saves the masks along with the caches. 
While stitching, \name{} restores the caches and masks of two latent states, extracts the required part from each of the latent states, and combines them into a single latent state.
\name{} will combine the two masks in case they do not perfectly align. 
Typically, a good prompt similarity guarantees that such misalignment is minor. 
After taking the remaining denoising steps, this stitching will not introduce significant quality loss.
This new latent state is then taken as input to the video diffusion model.

\begin{figure}
    \centering
    \includegraphics[width=1\linewidth]{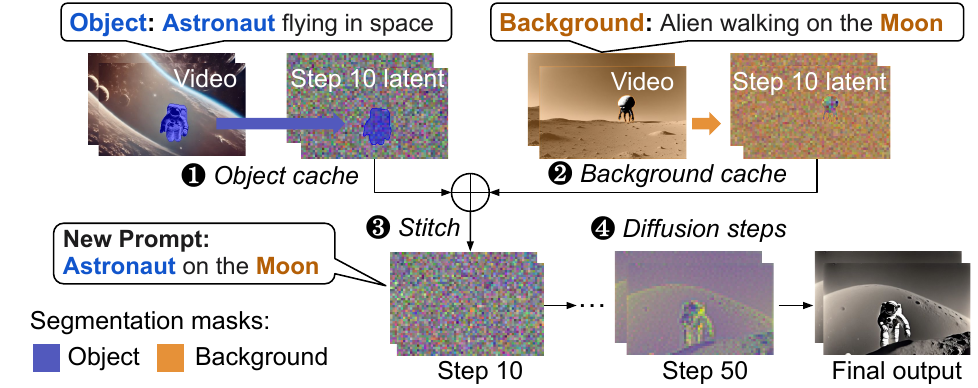}
    \caption{Illustration of cache reconstruction.}
    \label{fig:reconstruction_example}
\end{figure}

\Cref{fig:reconstruction_example} demonstrates the workflow cache reconstruction. 
There is a new request for a video about \emph{astronaut} on the \emph{Moon}. 
The vector DB does not have a good match for the whole prompt, so it provides two caches:  ``\emph{astronaut} flying in space''~\circled{1} and ``alien walking \emph{on the Moon}''~\circled{2}.
The former provides the object and the latter provides the background component. 
Thus, \name{} extracts the astronaut part from the former cache, and the moon part from the later cache using the previously saved masks to stitch them into one state~\circled{3}. After stitching, the new request can proceed from the \nth{11} step~\circled{4}.
The ``alien'' part in the background cache will be replaced by pixels in the object cache at the same location. 
This minor misalignment does not lead to any observable blurred pixels or extra objects in the final output. 

\subsubsection{Cache Replacement} \label{subsec:cache_replacement}

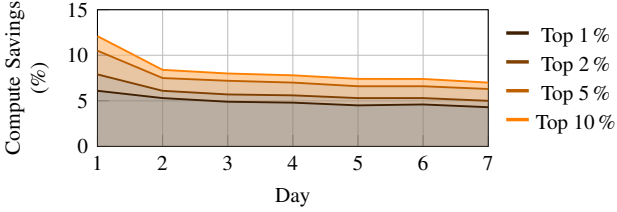
\begin{figure}[t]
    \centering
    \input{results/hit_decline}
    \caption{Compute savings from top 1\,\% -- 10\,\% cached prompts over time.}
    \label{fig:hit_over_time}
\end{figure}


A cache replacement policy aims to evict the least desirable cache entry in the cache upon a new insertion.
\nirvana{} introduces a replacement policy -- Least Computationally Beneficial and Frequently Used (LCBFU) for their image caching system. 
It selects the cache entry that has saved the least computations. 
Suppose the whole cache has $i$ cache entries, it uses $f_i \times step_i$ to determine the priority of i\textsuperscript{th} cache and evicts the cache with the least priority. 
However, we observe that video caching is in stark contrast to image caching.

\textbf{Observation 1: Non-uniform cache size.} The size of each cache entry (\ie latent of video) is not constant, as the number of extra frames varies by steps or prompt. 
As a result, replacing one cache entry may lead to different benefits in storage. 
In comparison, the existing image cache replacement policy LCBFU only captures the performance benefits since image caching only has uniform-sized cache entries.

\textbf{Observation 2: Decreasing reuse over time.} Recency is equally critical for the replacement policy. We conduct an experiment based on VidProM \cite{wang2024vidprommillionscalerealpromptgallery}, a real-world user trace dataset. We track the change of computation savings from the hottest prompts (\ie most frequently hit) over time. First, we select the prompts that contributes top 1--10\% computation savings to a single day. Then, we further collect their savings for the following days. 
The result is shown in \Cref{fig:hit_over_time}. 
We observe that the top 1\,\% prompts remain popular for a long
time. In contrast, top 10\,\% prompts exhibit a noticeable
decline in popularity over days.
LCBFU overlooks recency and tends to save caches frequently accessed in the past but not recently reused. 
\textbf{\cachename{}:} To evict caches that are less likely to be reused in the future, we propose a novel cache replacement policy named \textit{Least Reused Benefit Unit} (\cachename{}), which takes both memory capacity and recency into consideration based on the observations mentioned above. \cachename{} treats individual step's cache as separated cache entries and evicts one of them will not influence other caches from the same request. We define the priority of i\textsuperscript{th} cache as:
\begin{equation} \label{eqa:replacement}
    \mathit{Priority}_{i} = \frac{f_i \times \mathit{step}_{i}}{\mathit{Capacity}_{i} \times \mathit{Duration}_{i}}
\end{equation}
where $f_i$ denotes the access frequency of the cache, $\mathit{step}_{i}$ denotes the number of steps that can be saved, $\mathit{Capacity}_{i}$ denotes the size of the cache entry (\ie number of frames), and $\mathit{Duration}_{i}$ represents the time elapsed since the cache was last accessed. Cache entry with the lowest $\mathit{Priority}$ will be evicted first. 
Specifically, $\mathit{Capacity}_{i}$ in \Cref{eqa:replacement} aims to normalize the benefit by its capacity; $\mathit{Duration}_{i}$ aims to evict caches that have not been reused for a long time, preventing stale cache entries that were accessed frequently in the past from staying in the cache system forever. 

\begin{figure}[t]
  \begin{center}
  \includegraphics[width=\linewidth]{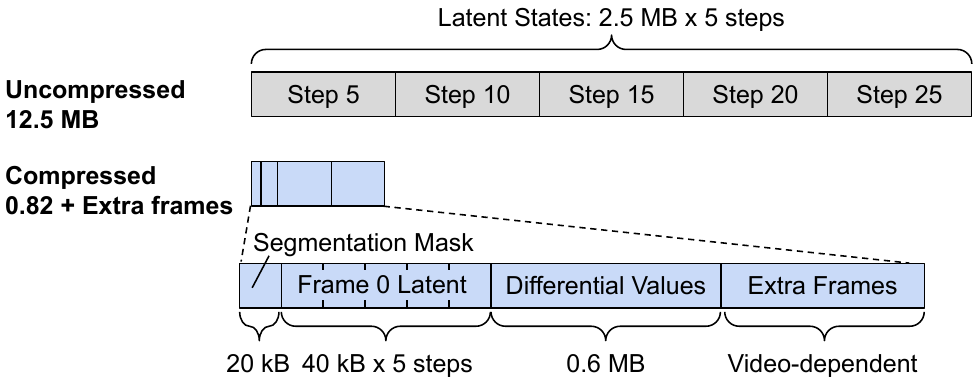}
  \end{center}
  \caption{\label{fig:memory-breakdown} Memory consumption breakdown for uncompressed and compressed cache. }
\end{figure}

\subsubsection{Cache Operations} \label{subsubsec:cache_operations}
The cache in \name{} takes two operations: 

\textbf{Insertion.} Unlike LCBFU, which only saves the cache upon a miss, the efficient latent state compression allows \cachename{} to save more states. 
When a prompt hits a cache (either whole prompt or decoupled) and the step it hits is $n$. 
If $n < 25$, \ie the highest step that can be cached, \cachename{} saves the additional key steps between $n$ and 25. 
If $n = 25$, then no extra states will be saved. 
This mechanism allows more later steps to be saved, providing more potential computation savings for future prompts that are similar to the one that was just processed. 

\textbf{Eviction.} 
\cachename{} keeps track of all the cached latent states. 
\cachename{} updates their duration each time when it is going to start evicting and then discards enough caches until the new caches can be inserted.
The eviction mechanism chooses the cache entry with the lowest $\mathit{Priority}$ according to \Cref{eqa:replacement}.
Because the latent states of the same prompt from different steps can be evicted separately, some states of certain steps may be evicted earlier, leaving a ``hole''. 
We take the same approach as \nirvana{}'s solution to the ``hole'' problem.
If an incoming prompt hits an already evicted step from an existing prompt, \name{} takes an earlier step instead. 


\subsection{Cache Entry Structure}
Putting the caching compression and hit rate optimization techniques together, 
\Cref{fig:memory-breakdown} shows each component in a compressed cache entry that maintains the \nth{5}, \nth{10}, \nth{15}, \nth{20}, and \nth{25} steps of latent states. 
Directly saving these 5 steps takes \SI{12.5}{\mega\byte} for each prompt. 
\name{}'s compression technique significantly reduces the space requirement. 
For each prompt, the cache entry maintains the first frame of each step, the differential value generated by the common key frames (\ie the key frame that exists for all 5 steps), and the extra frames as some states have more key frames than others. 
To enable the decoupled cache lookup, the segmentation mask takes \SI{40}{\kilo\byte} as it takes only 1 bit for each pixel.
In our evaluation, we find that the cache size can reach up to \SI{4.9}{\mega\byte}, which is still much less than uncompressed. 
See \Cref{subsubsec:compression_efficiency_quality} for detailed evaluations. 

%% file: results/diff_heatmap.tex
\begin{tikzpicture}
\begin{axis}[%
    height=4cm,
    font=\footnotesize,
    tick align=outside,
    xlabel near ticks,
    xmin=2.5, xmax=52.5,
    xtick={5,10,15,20,25,30,35,40,45,50},
    xtick style={draw=none},
    xlabel=Step,
    xlabel style={font=\footnotesize},
    xticklabel style={rotate=90},
    xlabel shift={-3pt},
    xticklabel shift={-4pt},
    ymin=2.5, ymax=52.5,     
    ytick={5,10,15,20,25,30,35,40,45,50},
    ytick style={draw=none},
    ylabel=Step,
    ylabel style={font=\footnotesize},
    ylabel shift={-3pt},
    yticklabel shift={-4pt},
    point meta min=0,
    point meta max=1,
    point meta=explicit,
    colorbar,
    colorbar horizontal,
    colorbar style={
        at={(0,1.42)},anchor=north west,
        width=2.4cm,
        height=0.2cm,
        xlabel={Similarity},
        xlabel shift={-3pt},
        xlabel style={font=\footnotesize},
        xtick style={draw=none,font=\footnotesize},
        xticklabel style={font=\footnotesize},
        ylabel style={anchor=south},
    },
    colormap={yellowblue}{color=(blue) color=(yellow!80!orange)},
    scale mode=scale uniformly,
]
 \addplot [
        matrix plot,
        mesh/cols=10,
        point meta=explicit,
        draw=white,
        line width=0.1mm,
] table [meta=similarity,col sep=comma] {results/csv/diff_similarity_matrix.csv};
\draw [gray, line width=0.4mm] (2.5,2.5) rectangle  (27.5,27.5);
\node[draw=none,align=center, text width=1cm] at (15,15) {Cached Steps};

\end{axis}
\end{tikzpicture}

\vspace{-2mm}

%% file: results/similarity_hit.tex
\begin{tikzpicture}
\begin{axis}[
style={font=\footnotesize},
xlabel={Similarity score},
ylabel={\% Prompts},
width=1\hsize,
scaled y ticks=false,
xtick pos=bottom,
xtick = {0.5,0.6,0.7,0.8,0.9},
ytick pos=left,
ytick = {0,5,10,15,20,25},
xmin=0.5,
xmax=0.95,
ymin=0,
ymax=25,
clip=false,
height=3.7cm,
grid=both,
legend columns=1,
legend cell align=left,
legend style={
cells={align=left},
anchor=north,
at={(0.35,1.65)},
draw=none,
fill=none,
column sep=0.5ex,
},
legend image code/.code={%
    \draw[#1, draw] (-0.1cm,0.05cm) -- (0.1cm,0.05cm);
}, 
]
\addplot+[
name path=A,
draw=black,
line width=0.25mm,
mark=none,
] table[col sep=comma, x=score, y=Full prompt] {results/csv/similarity_hit.csv};
\addplot+[
name path=B,
draw=blue,
line width=0.25mm,
mark=none
] table[col sep=comma, x=score, y=Object] {results/csv/similarity_hit.csv};
\addplot+[
name path=C,
draw=orange,
line width=0.25mm,
mark=none
] table[col sep=comma, x=score, y=Background] {results/csv/similarity_hit.csv};
\addplot+[draw=none,mark=none,name path=xaxis] coordinates 
 {(0.5, 0) (0.95,0)}; 
\addplot+[black!40!green, fill opacity=0.1] fill between[of=A and xaxis, soft clip={domain=0.65:0.95}];
\addplot+[black!40!green, fill opacity=0.1] fill between[of=B and xaxis, soft clip={domain=0.65:0.95}];
\addplot+[black!40!green, fill opacity=0.1] fill between[of=C and xaxis, soft clip={domain=0.65:0.95}];
\draw[black!40!green,dashed,line width=0.4mm] (0.65,0) -- (0.65,25);
\node[draw=none,color=black!40!green] at (0.7,22) {\normalsize Hit};
\legend{Whole Prompt (hit=78.9\,\%), Object (hit=92.7\,\%), Background (hit=95.9\,\%)};

\end{axis}
\end{tikzpicture}

\vspace{-2mm}

%% file: results/hit_decline.tex
\begin{tikzpicture}
\begin{axis}[
style={font=\footnotesize},
ylabel style={align=center, text width=3cm},
xlabel={Day},
ylabel={Compute Savings\\ (\%)},
width=0.8\hsize,
scaled y ticks=false,
xtick pos=bottom,
ytick pos=left,
xtick = {1,2,3,4,5,6,7},
xmin=1,
xmax=7,
ymin=0,
ymax=15,
height=3.4cm,
grid=both,
legend columns=1,
legend image code/.code={%
    \draw[#1, draw, line width=1.3pt] (-0.15cm,0.05cm) -- (0.15cm,0.05cm);
}, 
legend style={fill=none,nodes={scale=1},draw=none,anchor=north west,at={(1.02,0.95)}},
]
\addplot+[
name path=top 1p,
draw=orange!25!black,
line width=0.25mm,
mark=none,
] table[col sep=comma, x=day, y=top 1p] {results/csv/hit_decline.csv};
\addplot+[
name path=top 2p,
draw=orange!50!black,
line width=0.25mm,
mark=none,
] table[col sep=comma, x=day, y=top 2p] {results/csv/hit_decline.csv};
\addplot+[
name path=top 5p,
draw=orange!75!black,
line width=0.25mm,
mark=none,
] table[col sep=comma, x=day, y=top 5p] {results/csv/hit_decline.csv};
\addplot+[
name path=top 10p,
draw=orange!100!black,
line width=0.25mm,
mark=none,
] table[col sep=comma, x=day, y=top 10p] {results/csv/hit_decline.csv};
\addplot+[draw=none,mark=none,name path=xaxis] coordinates {(1,0) (7,0)}; 
\addplot+[orange!25!black!60!white, fill opacity=0.6] fill between[of=top 1p and xaxis, soft clip={domain=1:7}];
\addplot+[orange!50!black!60!white, fill opacity=0.6] fill between[of=top 2p and top 1p, soft clip={domain=1:7}];
\addplot+[orange!75!black!60!white, fill opacity=0.6] fill between[of=top 5p and top 2p, soft clip={domain=1:7}];
\addplot+[orange!100!black!60!white, fill opacity=0.6] fill between[of=top 10p and top 5p, soft clip={domain=1:7}];
\legend{Top 1\,\%, Top 2\,\%, Top 5\,\%, Top 10\,\%};

\end{axis}
\end{tikzpicture}
\vspace{-4mm}

%% file: sec/4_impl.tex
\section{Implementation of \name{}}
\label{sec:impl}


 \begin{figure}[t]
  \begin{subfigure}[b]{0.48\linewidth}
    \includegraphics[width=1\linewidth]{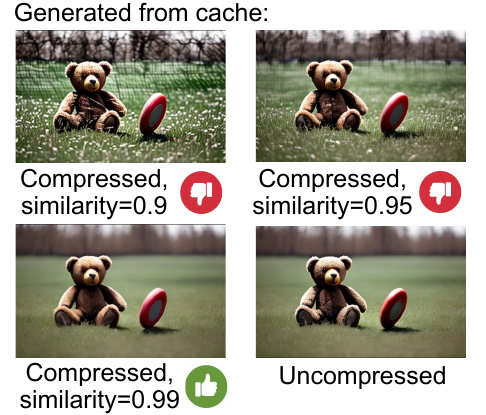}
    \caption{\label{subfig:compress_demo}}
  \end{subfigure}
  \hspace{2mm}
  \begin{subfigure}[b]{0.48\linewidth}
  \input{results/redundant_frames}
  \caption{\label{subfig:redundant_frames}}
  \end{subfigure}
  \caption{\label{fig:keyframes} 
  (a) Quality of video under different similarity levels in compression. 
  (b) Number of redundant latent frames among steps when the similarity threshold is 0.99.
  }
\end{figure}


We implement \name{} in PyTorch \cite{torch} with \SI{1.9}{\thousand} line of codes. 
The implementation details of components in \Cref{fig:arch} are the following:


\textbf{Diffusion Model.}
We use the industry-standard VideoCrafter2 \cite{videocrafter} as the text-to-video diffusion model, with the integration of  
FreeNoise \cite{freenoise} to generate 64-frame videos (4 seconds). 
We use xformers \cite{xFormers2022}, one of the most popular transformer frameworks, to help optimize the diffusion model as other projects \cite{podell2023sdxl, Bolya_2023_CVPR, Ye_Liu_Wu_Wu_2024, HiDiffusion}.
We follow the default configuration of VideoCrafter2, with video resolution of $320 \times 512 \times 4$,  $8 \times$ downscaled latent states for both height and width, and 50 denoising steps.

\textbf{Text extractor. }
We use Llama2-7B \cite{llama2} to extract the background and object descriptions from the input prompt. 
The model is based on float-16. 
We set the maximum generated length as the maximum number of tokens that the diffusion model can process. 
To alleviate the overhead brought by this add-on module, we apply Flash Attention \cite{dao2022flashattention, dao2023flashattention2} to speed up Llama execution.

\textbf{Vector database. }
We use Qdrant \cite{qdrant} as the vector DB for cache lookup.
\name{} has three sets of prompt embeddings: object, background, and whole prompt, where each set of embeddings is kept in one index table. 
The query results of the vector database are pointers to the cache entry of the video latent states. Thus, even though we save three times as many indices, the total size of caches remains the same. 
For example, \SI{700}{\thousand} unique prompts take \SI{1000}{\giga\byte} to store their caches but all of the three types of embeddings only take \SI{6}{\giga\byte}.
Upon a new prompt, \name{} save its background, object, and itself to the three index tables and set the value pointing to the same location in the storage. 
For the background similarity score and object similarity score, we choose the minimum value as the final score as this choice guarantees the quality of the generated video.


\textbf{Cache compressor. }
We compress the video latent states using the method in \Cref{subsec:compression}. 
\Cref{subfig:compress_demo} demonstrates the impact of different similarity thresholds (based on cosine similarity).
Here we showcase three thresholds: 0.9, 0.95, and 0.99, and compress the latent of the \nth{25} step.
The result indicates that a threshold of 0.99 is necessary to guarantee high quality; lower thresholds lead to noticeable noise such as undesired patterns. 
Therefore, we take 0.99 as the threshold in our implementation. 
\Cref{subfig:redundant_frames} shows the percentage of redundant frames of each step with the similarity threshold of 0.99. As the step of the cache increases, the number of redundant frames reduces, as later steps contain more details. 

\textbf{Video segmenter. }To achieve better segmentation results with given object and background prompts, we use LangSAM~\cite{langsam} to figure out the boundaries of objects accurately.
LangSAM is an open-source project that can perform instance segmentation and use text prompts to generate masks.
Because it only works for images, we treat the whole video as a sequence of images and segment them to get masks for each frame. 
Since the masks only take 1 bit per pixel, saving the whole video's masks is low-overhead but enables better stitching. 
Video segmentation is not on the critical path of \name{} as it happens after the video has been generated. Besides, LangSAM is relatively lightweight. 
Therefore, we use the CPU to generate the mask, by starting another process to run the LangSAM-based segmentation while \name{} processes new incoming prompts. 

%% file: results/redundant_frames.tex
\begin{tikzpicture}
\begin{axis}[
ybar,
bar width=3,
style={font=\footnotesize},
xlabel={Cached Step},
ylabel={\% Redundant Frames},
width=1\hsize,
scaled y ticks=false,
xtick pos=bottom,
xtick style={draw=none},
ylabel near ticks, 
ylabel shift={-5pt},
ytick pos=left,
ytick = {0,25,50,75,100},
xmin=0,
xmax=30,
xtick = {5,10,15,20,25},
ymin=0,
ymax=100,
height=3.5cm,
ymajorgrids=true,
legend columns=2,
legend cell align=left,
legend style={draw=none,anchor=north,at={(0.5,1.3)}},
legend image code/.code={%
    \draw[#1, draw] (0cm,-0.05cm) rectangle (0.2cm,0.1cm);
}, 
]
\addplot+[
    draw=blue!70!white!90!black,
    fill=blue!40!white!90!black,
    y filter/.expression={y/64*100},
] table[x=step num,y=redundant frames,col sep=comma] {results/csv/redundant_frames.csv};

\end{axis}
\end{tikzpicture}

%% file: sec/5_eval.tex
\section{Evaluation} 

\subsection{Evaluation Methodology}
\label{subsec:meth}

In this section, we discuss the methodology of our evaluation.


\textbf{Platform.}
We evaluate the \name{} on GCP machine type \machinetype{} that comes with an Nvidia A100 \SI{40}{\giga\byte} GPU, 12 vCPU cores, and \SI{85}{\giga\byte} main memory. We add an additional \disktype{} to store latent state caches. 
We compare the following video generation schemes:

\begin{itemize}[leftmargin=*,noitemsep,partopsep=2pt,topsep=2pt,parsep=0pt]
    \item \textbf{No Cache}: Use the original model to directly generate videos, without caching prompts. 
    \item \textbf{\nirvana{}-video}: Directly adapt \nirvana{} \cite{nirvana} to perform video caching, without cache compression, or decoupled object and background cache lookup. 
    \item \textbf{\name{} (this work)}: Include all optimization techniques form \Cref{sec:design}. 
\end{itemize}

\textbf{Dataset.} \label{subsec:dataset}
We evaluate \name{} on dataset VidProM \cite{wang2024vidprommillionscalerealpromptgallery}, which contains \SI{1.67}{\million} unique text-to-Video prompts from real-world users on Discord. Each prompt is associated with a timestamp. 
We select first \SI{700}{\thousand} prompts according to their timestamps for evaluation, which takes around \SI{1000}{\giga\byte} of caching data.
VidProM is similar to  DiffusionDB, a commonly used text-to-image dataset  \cite{diffusiondb} that is based on real-world users. However, VidProM has 40.6\,\% more semantically unique prompts, making cache reuse more challenging. 
Nonetheless, our evaluation of cache hit rate (\Cref{subsec:replacement_policy_result}) demonstrates that \name{} enables a high hit rate using our hit rate optimization strategies. 

\textbf{Cache Replacement Policies.} \label{subsec:replacement_policies}
We evaluate the following cache replacement policies for video caching. 

\begin{itemize}[leftmargin=*,noitemsep,partopsep=2pt,topsep=2pt,parsep=0pt]
    \item First In, First Out (\textbf{FIFO}): Replace cache entries in the same order they were inserted, where the oldest entry gets evicted first.  
    \item Least Recently Used (\textbf{LRU}): Replace the least recently used cache entry. 
    \item Least Computationally Beneficial and Frequently Used (\textbf{LCBFU}):
    The replacement policy in \nirvana{} \cite{nirvana} for image caching, as discussed in \Cref{subsec:cache_replacement}. We adapt this method to video caching. 
    \item Least Reused Benefit Unit (\textbf{\cachename{}}) in \textbf{this work}: the replacement policy of \name{} as introduced in \Cref{subsec:cache_replacement}.
\end{itemize}

\textbf{Video Quality Metrics.} 
\label{subsec:metric}
We compare the video generation quality using the following metrics that are commonly used in evaluating video quality.  
\begin{itemize}[leftmargin=*,noitemsep,partopsep=2pt,topsep=2pt,parsep=0pt]
    \item \textbf{FVD \cite{freenoise}} calculate the distance between the generated video and the original video in dataset \cite{unterthiner2019fvd}. A lower FVD indicates the two videos are more similar. Typically, a difference of less than 50 between two generated video sets means the two sets have no visual difference \cite{unterthiner2019fvd}. 
    \item \textbf{CLIP-Text \cite{wu2024bettermetrictexttovideogeneration, videobooth}} get the average value of the cosine similarity scores between prompt and each frame in video \cite{videobooth}. A high score indicates better alignment between the video and the prompt. 
    \item \textbf{CLIP-SIM \cite{freenoise, pmlr-v139-radford21a}} is generated by computing the average cosine similarity scores of the adjacent frames. A higher score indicates better quality \cite{radford2021learningtransferablevisualmodels,qiu2024freenoisetuningfreelongervideo}. 
\end{itemize}

\subsection{Evaluation Results} 

In this section, we evaluate the performance, generation quality, and cost of \name{}. 

\subsubsection{Performance Breakdown}

\begin{figure}
    \centering
    \includegraphics[width=1\linewidth]{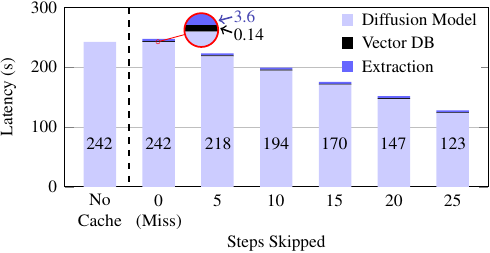}    
    \caption{Latency breakdown.}
    \label{fig:latency_breakdown}
\end{figure}

As discussed in \Cref{subsec:sys_overview}, each prompt undergoes three main operations in \name{}: extraction, vector DB lookup, and diffusion model. 
\Cref{fig:latency_breakdown} shows the latency breakdown (y-axis) when the prompt skips 0 (\ie a cache miss),  5, 10, 15, 20, and 25 steps. In addition, we show the generation latency when no cache is used. 
We track the latency breakdown of 100 prompts and present the average. 
As the diffusion model is the most time-consuming, the bottom stack takes the majority of latency, ranging from \SI{242}{\second} to \SI{123}{\second}. We find that the latency reduction of the diffusion model is proportional to the skipped steps. 
In comparison, vector DB takes an average of \SI{0.14}{\second} --- a latency visible only within the zoom-in circle. 
Extraction that uses Llama2-7B takes an average of \SI{3.6}{\second}. Nonetheless, this latency is only \SI{1.5}{\percent} of the diffusion model.

\subsubsection{Throughput Comparison} \label{subsec:throughput}

\begin{figure}
    \centering
    \input{results/throughput}
    \caption{Throughput of video generation.}
    \label{fig:throughput}
\end{figure}
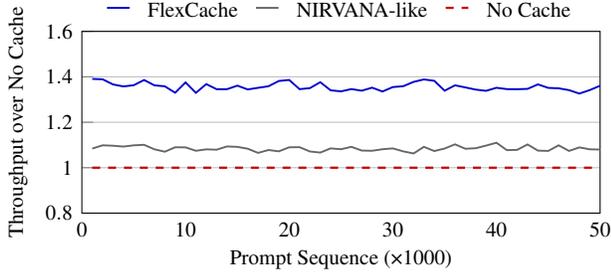

\textbf{Throughput over time.}
This experiment demonstrates the throughput of video generation. 
we first set up a capacity limit of \SI{1}{\tera\byte} and fill up the cache with prompts. 
Then, we evaluate \SI{50}{\thousand} prompts according to their timestamps and record the generation throughput and cache hit rate, where we take the average value of every \SI{1}{\thousand} prompts as one data point. 
\Cref{fig:throughput} shows the result, where the y-axis shows the throughput normalized to the ``No Cache'' baseline and the x-axis shows the prompt sequence. The result indicates that our system gets 1.26$\times$ better throughput than \nirvana{}-video on average. 

\textbf{Throughput Breakdown.}
Using the same setup as above, we break down the throughput improvements over the ``No Cache'' scheme, as \Cref{fig:throughput_breakdown} shows.
With latent caching alone, the improvement is 1.08$\times$. 
By applying cache compression (\Cref{subsec:compression}) on top, the improvement becomes 1.13$\times$.
Finally, integrating the cache hit rate optimizations in \name{}, including the \cachename{} replacement policy and decoupled object and background cache lookup improves the throughput to 1.36$\times$ of the ``No Cache'' scenario.


\begin{figure}[t]
\begin{minipage}[t]{0.38\linewidth}
    \centering
    \input{results/throughput_breakdown}
    \caption{Breakdown of throughput gain. }
    \label{fig:throughput_breakdown}
\end{minipage}
\hspace{0.2cm}
\begin{minipage}[t]{0.58\linewidth} 
    \centering
    \input{results/hit_step_dist}
    \caption{Distribution of number of steps skipped. }
    \label{fig:hit_step_dist}
\end{minipage}        
\end{figure}


\textbf{Steps Skipped}
We next evaluate the cache hit rate by taking the same prompt sequence and capacity as \Cref{subsec:throughput}. 
The y-axis in \Cref{fig:hit_step_dist} shows the percentage of prompts that skip 0, 5, 10, 15, 20, and 25 steps, where 0 step indicates a miss. 
Overall, the hit rate is \SI{98.4}{\percent}, with skipping 10 steps being the most common (\SI{35.4}{\percent}).


\subsubsection{Cache Replacement Comparison} \label{subsec:replacement_policy_result}

\begin{table}[t]
\setlength{\tabcolsep}{4pt}
    \caption{Comparison among cache replacement policies.} 
    \label{tab:cache_replacement}
    \centering
    \small
    \begin{tabular}{cccccc}
    \toprule
      Policy & Cache Size &  FIFO & LRU & LCBFU & \textbf{\cachename{}} \\
      \midrule
      \multirow{4}{*}{Hit Rate} & \SI{1}{\giga\byte} & 38\,\% & 68\,\% & 54\,\% & 86\,\% \\
      & \SI{10}{\giga\byte}  & 40\,\% & 84\,\% & 65\,\% & 91\,\% \\
      & \SI{100}{\giga\byte}  & 44\,\% & 94\,\% & 88\,\% & 95\,\% \\
      & \SI{1000}{\giga\byte} & 48\,\% & 98\,\% & 97\,\% & 98\,\%\\
      \midrule
      \multirow{4}{*}{\shortstack{Computation\\Savings}} & \SI{1}{\giga\byte} & 10\,\% & 15\,\% & 13\,\% & 18\,\%\\
      & \SI{10}{\giga\byte}  & 11\,\% & 19\,\% & 15\,\% & 21\,\% \\
      & \SI{100}{\giga\byte} & 14\,\% & 23\,\% & 20\,\% & 24\,\% \\
      & \SI{1000}{\giga\byte} & 17\,\% & 27\,\% & 26\,\% & 28\,\% \\
      \bottomrule
    \end{tabular}
\end{table}

\cachename{} is the default cache replacement policy in \name{}. We perform a sensitivity study by replacing \cachename{} with other cache replacement policies in \Cref{subsec:replacement_policies}.
We first initialize the cache to reach its size limit and then evaluate \SI{50}{\thousand} prompts. 
\Cref{tab:cache_replacement} shows the hit rate and compute savings under cache sizes of \SI{1}{\giga\byte}, \SI{10}{\giga\byte}, \SI{100}{\giga\byte}, and \SI{1000}{\giga\byte}.
Note that the computation savings measure the reduced diffusion model computations to reflect the correlation between the overall hit rate and actual skipped steps from each cache. 
As the cache capacity increases, the hit rate and computation savings also increase. 
Overall, \cachename{} has the highest computation savings among the four replacement policies.

\begin{figure}[t]
\begin{minipage}[t]{0.48\linewidth} 
    \centering
    \input{results/extra_frames}
    \caption{Size of latent of extra frames.}
    \label{fig:extra_frames}
\end{minipage}
\hspace{1mm}
\begin{minipage}[t]{0.48\linewidth} 
    \centering
    \input{results/compression_ratio}
    \caption{Distribution of compression ratio.}
    \label{fig:compression_ratio}
\end{minipage}
\end{figure}

\subsubsection{Compression Efficiency and Quality}
 \label{subsubsec:compression_efficiency_quality}

\begin{figure}[t]
\begin{minipage}[b]{0.52\linewidth} 
    \centering
    \input{results/decompress_similarity}
    \caption{Quality of intra- and inter-step compression.} 
    \label{fig:decompress_similarity}
\end{minipage}
\hspace{1mm}
\begin{minipage}[b]{0.45\linewidth} 
    \centering
    \input{results/cost_breakdown}
    \caption{Cost of video generation.}
    \label{fig:cost_breakdown}
\end{minipage}
\end{figure}


We conduct an experiment with latent states from \SI{5}{\thousand} prompts to find the distribution of extra frames for different steps.
\Cref{fig:extra_frames} demonstrates how the extra key frames change with steps. 
States from the \nth{5} and \nth{10} steps have no extra frames, while those later steps can have more frames. 
With the number of steps increasing, the latent state has a higher chance of having extra frames. \Cref{fig:compression_ratio} shows the distribution of the compression ratio for these two compression techniques. We find that 31\,\% caches can achieve the highest $16\times$ compression ratio (when there are no extra frames); 
the worst-case 1\,\% caches can get the least 2.53$\times$ compression ratio.
Both compression techniques have similar compression ratio distributions as the reduction of compression ratio is due to the extra frames that are saved separated. 
On average, \name{} reduces the cache size by 6.7$\times$.


To evaluate the quality of compression, we first apply the intra-step compression and then apply the inter-step compression on top. 
\Cref{fig:decompress_similarity} shows the cosine similarity between the original, uncompressed latent state and the decompressed latent state. Both these two compression mechanisms only introduce little loss to the original cache.
Overall, the similarity scores remain high ($>0.995$), indicating a negligible difference in video quality.

\subsubsection{Generation Quality}

We compare the quality of generation using metrics in \Cref{subsec:metric}. 
We conduct an experiment with \SI{2}{\thousand} videos to get all of those metrics. 
Similar to the approach in FreeNoise \cite{freenoise}, we split the generated videos into multiple smaller videos to align frames in the generated with those in the reference video.
\Cref{tab:quality} presents the quality metrics of \name{}. In this table, we can find that all of these three metrics are highly acceptable. The FVD difference between \name{} and the No Cache version is less than 50, which means such a difference in the quality of the generated videos that is hard to perceive by humans \cite{unterthiner2019fvd}. For the CLIP-Text and CLIP-SIM, there is also only a negligible difference as prior works have shown \cite{wang2025zola, DynamiCrafter}. 
 
\begin{table}[t]
    \caption{Quality comparison. }
    \label{tab:quality}
    \centering 
    \small
    \begin{tabular}{lccc}
    \toprule
         & FVD  & CLIP-Text & CLIP-SIM \\
    \midrule
     No Cache & 168 & 0.25  & 0.94   \\
     \name{} & 192 & 0.24  & 0.92  \\
     \nirvana{}-video & 172 & 0.24 & 0.94 \\
     \bottomrule
    \end{tabular}
\end{table}

\subsubsection{Cost Savings}

As our evaluation is based on GCP, we follow their pricing to estimate the cost savings from adopting \name{}. 
The machine type \machinetype{} costs \gpucost{} and the storage \disktype{} for keeping video latent caches costs \storagecost{} every month (as of December 2024) \cite{gcp-pricing}.
We calculate the per-video cost when \SI{1}{\tera\byte} of storage is used, as shown in \Cref{fig:cost_breakdown}.
Directly applying \nirvana{}'s caching scheme to video only yields 4.8\,\% savings as compared to No Cache, as the additional storage cost offsets the savings. 
In comparison, \name{} has significant cost savings of 31\,\%. 

%% file: results/throughput.tex
\begin{tikzpicture}
\begin{axis}[
style={font=\footnotesize},
xlabel={Prompt Sequence (\texttimes 1000)},
ylabel=Throughput over No Cache,
width=1\hsize,
scaled y ticks=false,
xtick pos=bottom,
xtick style={draw=none},
xlabel shift ={-2pt},
ylabel near ticks, 
ytick pos=left,
xmin=0,
xmax=50,
ymin=0.8,
ymax=1.6,
height=4cm,
ymajorgrids=true,
legend columns=3,
legend cell align=left,
legend style={draw=none,fill=none,anchor=north,at={(0.5,1.22)},       
column sep=1ex,
},
legend image code/.code={%
    \draw[#1, draw] (-0.15cm,0.05cm) -- (0.15cm,0.05cm);
}
]
\addplot+[
    draw=blue!80!black,
    mark=none,
    line width=0.25mm,
] table[x expr=\coordindex+1, y=throughput,col sep=comma] {results/csv/throughput/CachedVid.csv};
\addplot+[
    draw=white!40!black,
    mark=none,
    line width=0.25mm,
] table[x expr=\coordindex+1, y=throughput,col sep=comma] {results/csv/throughput/nirvana.csv};
\addplot+[
    draw=red!80!black, 
    dashed, mark=none,
    line width=0.3mm,
] coordinates {(1,1) (50,1)};
\legend{\name{}, \nirvana{}-like, No Cache}

\end{axis}

\end{tikzpicture}
\vspace{-3mm}

%% file: results/throughput_breakdown.tex
\begin{tikzpicture}
\begin{axis}[
    style={font=\footnotesize},
    ybar stacked,
    ylabel shift={-2pt},
    bar width=15pt,
    height=3.4cm,
    width=1\hsize,
    ymin=1,
    ymax=1.4,
    x tick label style={align=center, text width=.8cm}, 
    xticklabels={},
    xlabel shift={-5pt},
    xlabel={\name{}},
    ylabel style={align=center}, 
    ylabel={Throughput over\\No Cache},
    xtick style={draw=none},
    ymajorgrids=true,
    reverse legend,
    legend columns=1,
    legend cell align=left,
    legend style={
        cells={align=left},
        anchor=north,
        at={(0.3,1.75)},
        draw=none,
        fill=none,
        column sep=0.5ex,
    },
    legend image code/.code={%
        \draw[#1,draw] (0cm,-0.07cm) rectangle (0.17cm,0.12cm);
    }
]

\addplot [draw=blue!30!white!70!black,fill=blue!20!white] 
table[col sep=comma, y=nirvana, meta=case, x expr=\coordindex] {results/csv/throughput_breakdown.csv};
\addlegendentry{Latent Caching}
\addplot [draw=blue!50!white!70!black,fill=blue!45!white] 
table[col sep=comma, y=segmentation, meta=case, x expr=\coordindex] {results/csv/throughput_breakdown.csv};
\addlegendentry{Cache Compression \S\ref{subsec:compression}}
\addplot [draw=blue!80!white!70!black,fill=blue!70!white]
table[col sep=comma, y=cache design, meta=case, x expr=\coordindex] {results/csv/throughput_breakdown.csv};
\addlegendentry{Cache Hit Rate Opt. \S\ref{subsec:hit_opt}}

\end{axis}
\end{tikzpicture}


%% file: results/hit_step_dist.tex
\begin{tikzpicture}
\begin{axis}[
ybar stacked,/pgf/bar shift=0pt,
bar width=.8,
style={font=\footnotesize},
xlabel={Steps Skipped},
ylabel={\% Prompts},
ylabel shift={-2pt},
width=1\hsize,
scaled y ticks=false,
xtick pos=bottom,
xtick style={draw=none},
ylabel near ticks, 
xlabel shift={-2pt},
ytick pos=left,
xmin=-1,
xmax=6,
xtick = {0,1,2,3,4,5},
xticklabels={0, 5, 10, 15, 20, 25},
ymin=0,
ymax=40,
height=3.3cm,
ymajorgrids=true,
legend columns=1,
legend cell align=left,
column sep=0.5ex,
legend style={draw=none,fill=none,anchor=north,at={(0.5,1.7)}},
legend image code/.code={%
    \draw[#1, draw] (0cm,-0.05cm)
    rectangle (0.2cm,0.1cm);
}, 
]
\addplot+[
    draw=green!70!black,
    fill=green!70!black!50!white,
] table[meta=step, x expr=\coordindex, y=whole prompt, col sep=comma] {results/csv/hit_step.csv};
\addplot+[
    draw=green!40!black,
    fill=green!40!black!50!white,
] table[meta=step, x expr=\coordindex, y=decouple, col sep=comma] {results/csv/hit_step.csv};
\addplot+[
    draw=red!80!black,
    fill=red!80!black!50!white,
] table[meta=step,x expr=\coordindex, y=miss,col sep=comma] {results/csv/hit_step.csv};
\legend{Hit whole prompt, Hit decoupled, Miss}

\node[draw=none, color=red!80!black] at (-0.1, 6) {1.6\,\%};


\end{axis}

\end{tikzpicture}

%% file: results/extra_frames.tex
\begin{tikzpicture}
\begin{axis}[
style={font=\footnotesize},
xlabel={Latent of extra frames (kB)},
ylabel={Percentile (\%)},
ylabel shift={-3pt},
width=1\hsize,
scaled y ticks=false,
xtick pos=bottom,
xtick = {0,1000,2000},
ytick pos=left,
height=3.5cm,
grid=both,
legend columns=2,
legend cell align=left,
legend style={
cells={align=left},
anchor=north,
at={(0.4,1.45)},
draw=none,
fill=none,
column sep=1ex,
},
legend image code/.code={%
    \draw[#1, draw] (-0.15cm,0.05cm) -- (0.15cm,0.05cm);
}, 
]
\addplot+[
dashed,
draw=gray!80!black,
line width=0.5mm,
mark=none
] table[col sep=comma, x=size, y=percentile] 
{results/csv/compression_size/step10_extra.csv};
\addplot+[
draw=red,
line width=0.25mm,
mark=none
] table[col sep=comma, x=size, y=percentile] 
{results/csv/compression_size/step15_extra.csv};
\addplot+[
draw=orange,
line width=0.25mm,
mark=none
] table[col sep=comma, x=size, y=percentile] 
{results/csv/compression_size/step20_extra.csv};
\addplot+[
draw=blue,
line width=0.25mm,
mark=none
] table[col sep=comma, x=size, y=percentile] 
{results/csv/compression_size/step25_extra.csv};
\legend{Step 5/10, Step 15, Step 20, Step 25};
\draw[dashed, draw=gray!80!black, line width=0.5mm] (0,0) -- (0,100);
\draw[->, line width=0.3mm] (450, 20) -- (40,20);
\node[draw=none, align=left, text width=2.2cm] at (1500,20) {Step 5/10: no\\extra frames};

\end{axis}
\end{tikzpicture}
\vspace{-2mm}

%% file: results/compression_ratio.tex
\pgfplotsset{
    inter_legend/.style={
        legend image code/.code={%
            \path[#1,fill=blue!20!white!90!black, fill opacity=0.5](0cm,-.1cm)rectangle(.3cm,.1cm);
        }
    }
}

\pgfplotsset{
    intra_legend/.style={
        legend image code/.code={%
            \path[#1,fill=blue!60!white!90!black, fill opacity=0.5](0cm,-.1cm)rectangle(.3cm,.1cm);
        }
    }
}

\begin{tikzpicture}
\begin{axis}[
style={font=\footnotesize},
ylabel style={align=center},
ylabel={Compression Ratio},
xlabel={Percentile (\%)},
width=1\hsize,
scaled y ticks=false,
xtick pos=bottom,
ytick pos=left,
ytick={0,5,10,15,20},
yticklabels = {0,5:1,10:1,15:1,20:1},
ylabel shift={-3pt},
xmin=0,
xmax=100,
ymin=0,
ymax=22,
clip=false,
height=3.5cm,
grid=both,
legend columns=1,
legend cell align=left,
legend style={
cells={align=left},
anchor=north,
at={(0.4,1.45)},
draw=none,
fill=none,
},
]
\addplot+[
intra_legend,
name path=intra_inter,
draw=black,
mark=none,
y filter/.expression={1/y},
x filter/.expression={100-x},
] table[col sep=comma, x=percentile, y=intra + inter compression ratio] {results/csv/compression_size/compression_ratio.csv};
\addplot+[
inter_legend,
name path=intra,
draw=black,
mark=none,
y filter/.expression={1/y},
x filter/.expression={100-x},
] table[col sep=comma, x=percentile, y=intra compression ratio] {results/csv/compression_size/compression_ratio.csv};
\addplot+[draw=none,mark=none,name path=xaxis] coordinates 
 {(0,0) (100,0)}; 
\addplot+[blue!60!white!90!black, fill opacity=0.6] fill between[of=intra and xaxis];
\addlegendentry{Intra-step compression};
\addplot+[blue!20!white!90!black, fill opacity=0.6] fill between[of=intra_inter and intra];
\addlegendentry{Inter-step compression};


\end{axis}

\end{tikzpicture}
\vspace{-2mm}

%% file: results/decompress_similarity.tex
\begin{tikzpicture}
\begin{axis}[
ybar,
bar width=1.25,
style={font=\footnotesize},
xlabel={Step},
ylabel={Similarity},
width=1\hsize,
scaled y ticks=false,
xtick pos=bottom,
xtick style={draw=none},
xtick = {5,10,15,20,25},
xlabel shift={-3pt},
xticklabel shift={-4pt},
ytick pos=left,
ytick = {0.97,0.98,0.99,1},
ymin=0.97,
ymax=1,
height=3.4cm,
ymajorgrids=true,
legend columns=1,
column sep=0.5ex,
legend cell align=left,
legend style={draw=none,anchor=south,at={(0.5,1.0)}},
legend image code/.code={%
    \draw[#1, draw] (0cm,-0.05cm) rectangle (0.2cm,0.1cm);
}, 
]
\addplot+[
    draw=blue!60!white!90!black,
    fill=blue!60!white!90!black,
    fill opacity=1,
] table[x=step,y=intra,col sep=comma] {results/csv/decompress_similarity.csv};
\addlegendentry{Intra-step};
\addplot+[
    draw=blue!20!white!90!black,
    fill=blue!20!white!90!black,
    fill opacity=1,
] table[x=step,y=intra+inter,col sep=comma] {results/csv/decompress_similarity.csv};
\addlegendentry{Intra- and inter-step};

\end{axis}
\end{tikzpicture}
\vspace{-2mm}

%% file: results/cost_breakdown.tex
\begin{tikzpicture}
\begin{axis}[
    style={font=\footnotesize},
    ybar stacked,
    bar width=8pt,
    clip=false,
    height=3cm,
    width=1\hsize,
    xmin=-0.3,
    xmax=2.3,
    ymin=0.15,
    ymax=0.26,
    xtick = {0,1,2},
    x tick label style={align=center}, 
    xticklabels = {No Cache,\nirvana{}\\-video,\name{}},
    ylabel={Cost per Video (\$)},
    xtick style={draw=none},
    xticklabel style={rotate=90},
    ymajorgrids=true,
    legend columns=2,
    legend cell align=left,
    legend style={
        cells={align=left},
        anchor=south,
        at={(0.42,1.02)},
        draw=none,
        fill=none,
        column sep=0.5ex,
    },
    legend image code/.code={%
        \draw[#1] (0cm,-0.07cm) rectangle (0.17cm,0.12cm);
    }
]

\addplot [draw=orange!60!white,fill=orange!40!white] table[col sep=comma, y=gpu, meta=system, x expr=\coordindex] {results/csv/cost_breakdown.csv};
\addlegendentry{GPU server}
\addplot [draw=orange,fill=orange!80!white] table[col sep=comma, y=disk, meta=system, x expr=\coordindex] {results/csv/cost_breakdown.csv};
\addlegendentry{Storage}

\end{axis}

\end{tikzpicture}
\vspace{-5mm}

%% file: sec/7_disscussion.tex
\section{Discussions and Related Works}

\textbf{Diffusion Model Optimizations.}
There are multiple researches that focus on improving the performance of diffusion models. DeepCache \cite{deepcache}, Block Caching \cite{block_cache}, and Learning-to-Cache \cite{ma2024learningtocacheacceleratingdiffusiontransformer} discover that the outputs of layers among multiple steps are similar, so they can cache the output from the last step and reuse it in the current step. FISEdit \cite{fisedit} finds that users tend to change only a little in the same session. It saves all the intermediate outputs from all model layers and steps. When the next request in the same session arrives, it attempts to reuse the existing outputs. 
These proposals can speed up diffusion models but are orthogonal to \name{} as they do not change the number of denoising steps.
However, it is possible to integrate these optimizations into \name{} to further reduce the video generation time.  

\textbf{Approximate Caching for image diffusion models.}
\nirvana{} \cite{nirvana} is the state-of-the-art approximate caching system for image diffusion models. While it is good in image generation, we have shown that it is not efficient in video generation.

\textbf{Support for other models.}
As far as we know, \name{} is the first approximate caching system for text-to-video diffusion models.
For evaluation, we integrated VideoCrafter2~\cite{videocrafter} as the diffusion model. Other diffusion models can be supported. 
Moreover, other diffusion-based generation tasks \cite{svd, make_a_video, DynamiCrafter, Chen_2024_CVPR} can also leverage our caching scheme.

\textbf{Creativity of generation.}
Since \name{} reuses the cache from the previous request and achieves a relatively high hit rate, the generated videos may tend to be less diverse over time. To alleviate this homogenization, \name{} deploy the same strategy as \nirvana{} \cite{nirvana} that changes the denoising seed after retrieval. 

%% file: sec/8_conclusions.tex
\section{Conclusions}
In this paper, we introduce the motivation and design of \name{} which further optimizes the approximate cache method for video generation. We compress the cache size to accommodate more caches under a fixed storage. Then we decouple the background and object for more efficient cache lookup, achieving higher hit rate and more computation savings.  We also provide a cache replacement policy tailored to the two designs above. Finally, the evaluation shows that our system can gain 1.26$\times$ throughput and 25\,\% cost savings over \nirvana{}'s approximate caching system.